\documentclass[%
 reprint,
 amsmath,amssymb,
 aps,
]{revtex4-2}

\usepackage{graphicx}
\usepackage{dcolumn}
\usepackage{bm}

\begin{document}

\preprint{APS/123-QED}

\title{Is it the truth?  Apparent ‘fake’ neutrino mixture due to seesaw mechanism responsible for oscillation}

\author{Xiao-Yan Wang}
 \altaffiliation[ ]{Physics Department, Sichuan University of Arts and Science.}
 \email{wangxy20220621@163.com}

\date{\today}

\begin{abstract}
This paper assumes that neutrino flavor conversion is induced by right-handed neutrino mixture via seesaw mechanism, which leads to apparent ‘fake’ neutrino mixture with neutrino mass eigenstate consistent with flavor state of left-handed neutrino rather than mixture of flavor state. The transition probability between right-handed neutrinos due to mixture can be explained well by boson intermediating flavor flip interaction between right-handed neutrinos and neutrino oscillation can be considered macro phenomenon before all flavor flip interactions arrive at balance.

\end{abstract}

\maketitle


\section{	Introduction}

Neutrino oscillation\cite{LSND,SK1998,DayaBay2012} can be described perfectly by neutrino mixture\cite{Bimaximal} which is due to the inconsistency of flavor state with mass eigenstate, while the mass eigenstates haven't been found yet so far. The possibility of neutrino oscillation without mass eigenstate have been studied in Ref.\cite{withoutweakeigenstate}. We find that it can be avoided if neutrino flavor conversion is induced  by right-handed neutrino mixture via seesaw mechanism\cite{Seesaw}, which is elaborated in this paper. 

The influence of right-handed neutrino on neutrino mixture have been studied in some articles\cite{cpphase,cp,cpbreak}, which assumed mixture existing in both left-handed and right-handed neutrino sector and part of properties of neutrino oscillation originated from right-handed neutrino mixture. In this paper, we furthermore assign that only right-handed neutrino mixture exists just like quark mixture only existing in down quark sector. Then dramatical apparent  ‘fake’ neutrino mixture will occur due to right-handed neutrino mixture by seesaw mechanism Eq. (1)
\begin{eqnarray}
m_\nu
&=&-\frac{m_Dm_D^T}{U_N^*M_N^d U_N^\dag}=-\frac{U_N^T m_Dm_D^TU_N}{M^d_N}\nonumber \\
&=&U^*\left(\begin{matrix}m_e&&\\&m_\mu&\\&&m_\tau\\\end{matrix}\right)U^\dag 
\end{eqnarray}
\begin{eqnarray}
M_N^d&=&\left(\begin{matrix}M_{N1}&&\\&M_{N2}&\\&&M_{N3}\\\end{matrix}\right) \nonumber
\end{eqnarray}
where $m_\nu$ is neutrino mass mixing matrix and $U$ is mixing matrix, $m_e,m_\mu,m_\tau$ are eigenvalues of neutrino mass eigenstates which are consistent with neutrino flavor states,
$m_D$ is neutrino Dirac mass matrix, $M_N^d$ is diagonal mass matrix of right-handed neutrino, $U_N$ is right-handed neutrino mixing matrix and $M_{N1}, M_{N2}, M_{N3}$ are eigenvalues of right-handed neutrino mass eigenstates. Obviously neutrino mixing is the conjugate transpose matrix of right-handed neutrino mixture $U=U_N^\dag$. And neutrino mixture is 'fake' which is actually right-handed neutrino mixture, and neutrino flavor does not change with time and distance by itself. Seesaw mechanism requires neutrino mass eigenstate to be superposition of left-handed state $ \nu_{L\alpha} \ (\alpha=e,\mu,\tau )$ and right-handed state $ N_R , -\frac{M_N}{m_D}\nu_L+N_R$. Without left-handed neutrino (denoted neutrino) mixing, neutrino mass eigenstate is consistent with neutrino flavor rather than flavor mixture. And for the same reason, mass eigenstate is not consistent with right-handed neutrino flavor but flavor mixture.

For the formalism of neutrino mixture, oscillation probability is given by Eq. (2)
\begin{eqnarray}
P\left(\nu_\alpha\rightarrow\nu_\beta\right)&=&{sin}^2\left(2\theta_{ij}\right){sin}^2\left(\frac{t}{L_0}\right) \\
L_0&=&\frac{4E}{\mathrm{\Delta}m_{ij}^2} \ \ ( \alpha, \beta= e,\mu, \tau) \nonumber   
\end{eqnarray}
where $\Delta m_{ij}^2=m_i^2-m_j^2 \ (i,j=1,2,3)$ is mass square difference of mass eigenstates which is flavor mixture. In this paper, without neutrino mixture, neutrino flavor conversion is induced by right-handed neutrino mixture via seesaw mechanism which can be considered as neutrino oscillation caused by apparent 'fake' mixture as shown in Fig. 1. (For simplicity, we assume no flavor conversion in seesaw and
right-handed neutrino mixture being a delayed mixture as shown in Fig. 1.)
\begin{figure}[!h]
\includegraphics[width=0.4\textwidth]{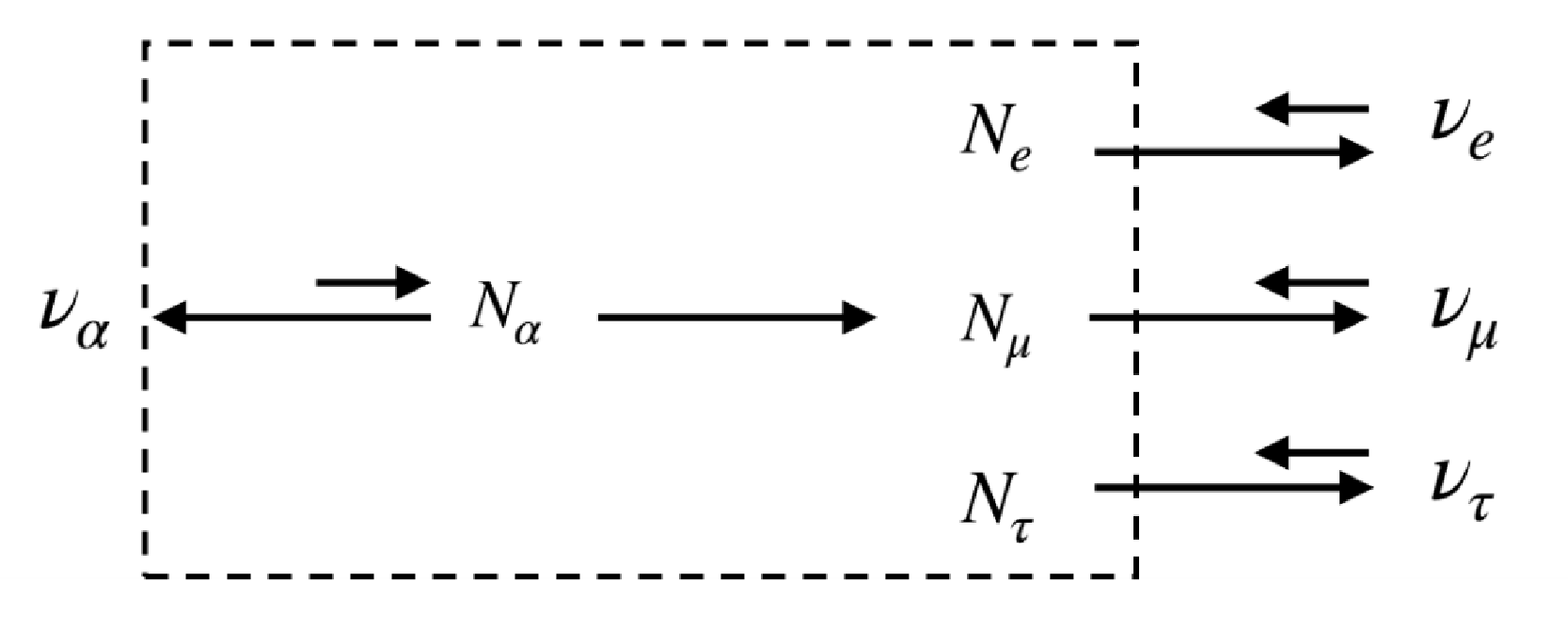}
\caption{ Apparent 'fake' neutrino mixture ( $\alpha= e,\mu, \tau$).}
\end{figure}
Thus oscillation probability formula Eq. (2) is still suitable for our formalism, except that the mass eigenstate should be consistent with flavor state and mass square difference should be $\Delta m_{ij}^2=m_i^2-m_j^2\ ( i,j= e,\mu, \tau)$.  
 
      Neutrino mass eigenstate being superposition of left-handed state and right-handed state $  -\frac{M_N}{m_D}\nu_{L\alpha}+N_R$ means that during flight, neutrino jumps between left-handed state $ \nu_{L\alpha} $ and right-handed state $ N_R$,  $\nu_{L\alpha}\rightleftharpoons N_R $. For the same reason, the assumed  right-handed neutrino mixture $ N_\alpha=\mathrm{\Sigma}_iN_{i} \ (\alpha=e,\mu,\tau;i=1,2,3) $ (where $N_\alpha$ is flavor state of right-handed neutrino, $N_{i}$ is  mass eigenstate of right-handed neutrino) means right-handed neutrino flavor conversion $N_\alpha\rightleftharpoons N_\beta $ after one flavor $N_\alpha$ have been produced in seesaw. Then neutrino flavor conversion will occur by chain of transition $ \nu_\alpha\rightarrow N_\alpha\rightarrow N_\beta\rightarrow\nu_\beta(\alpha,\beta=e,\mu,\tau)$, where we assume no flavor conversion in seesaw mechanism i.e. $ \nu_\alpha\rightarrow N_\alpha,N_\beta\rightarrow\nu_\beta $. For the mass of right-handed neutrino is very large, we assume all transitions in the chain  $ \nu_\alpha\rightarrow N_\alpha\rightarrow N_\beta\rightarrow\nu_\beta \ (\alpha,\beta=e,\mu,\tau) $ take place at the same position. The probability of neutrino flavor conversion $ \nu_\alpha\rightarrow N_\alpha\rightarrow N_\beta\rightarrow\nu_\beta $ can be evaluated by the formula $ P\left(v_\alpha\rightarrow N_\alpha\right)P\left(N_\alpha\rightarrow N_\beta\right)P\left(N_\beta\rightarrow\nu_\beta\right)$, where $P\left(v_\alpha\rightarrow N_\alpha\right) $ is transition probability of $ v_\alpha\rightarrow N_\alpha $ in seesaw mechanism, $ P\left(N_\alpha\rightarrow N_\beta\right) $ is transition probability  of $ N_\alpha\rightarrow N_\beta $ at one point due to right-handed neutrino mixture and $ P\left(N_\beta\rightarrow\nu_\beta\right) $ is transition probability of $N_\beta\rightarrow\nu_\beta $ in seesaw mechanism. According to the magnitude of $ P\left(N_\beta\rightarrow N_\beta\right) $ we will predict the possible origin of right-handed neutrino mixture.
       
       The paper is organized as follows: in Sec. 2, we elaborate the whole formalism for neutrino oscillation generated by 'fake' neutrino mixture, and we will evaluate the mass of right-handed neutrino according to seesaw relation and transition probability in seesaw in Part A, elaborate 'fake' neutrino mixture due to seesaw mechanism and evaluate flavor conversion probability in Part B, and  evaluate right-handed neutrino flavor conversion probability due to mixture and predict possible mechanism in Part C; in Sec. 3,  we will give a conclusion.

\section{	formalism}
This paper assumes that neutrino is Majorana particle which mass is given by seesaw mechanism and there exist only right-handed neutrino mixture which induces neutrino flavor conversion and neutrino apparent 'fake' mixture by seesaw mechanism which can depicts neutrino oscillation probability just like real neutrino mixture.

\subsection{\label{sec:level2}Neutrino mass and transition probability in seesaw}
    Introducing large mass Majorana right-handed neutrinos into the standard model (SM) and the Hamiltonian for neutrino mass is
\begin{eqnarray}
-m_D\overline{\nu_L}N_R-\frac{1}{2}N_R^TM_NN_R^C+h.c.
\end{eqnarray}
where  left-handed neutrino Majorana mass is assumed to be zero, $M_N$ is right-handed neutrino mass mixing matrix. The neutrino mass matrix is
\begin{eqnarray}
 -\left({\overline{\nu}}_L\overline{N_R}\right)\left(\begin{matrix}0&m_D\\m_D^T&M_N\\\end{matrix}\right) 
 \left(\begin{matrix}\nu_L \\ N_R^c\end{matrix}\right)
 \end{eqnarray} 
For Majorana particle, $\nu_L\equiv \nu_L^c,N_R\equiv N_R^c $, and effective mass of neutrino is
\begin{eqnarray}
  m_\nu\approx\frac{-m_Dm_D^T}{M_N}  .
 \end{eqnarray} 
Eq. (5) is called seesaw relation. And the neutrino mass eigenstate which diagonalize neutrino mass matrix is superposition of left-handed state and  right-handed state, $ -\frac{M_N}{m_D}\nu_{L\alpha}+N_R \ (\alpha=e,\mu,\tau)$. For lepton number of right-handed neutrino is zero, the number of right-handed neutrino is allowed a lot in the standard model. For the convenience and without loss of generality, we assume that there are three right-handed neutrinos $ N_e,N_\mu, N_\tau $ corresponding to three left-handed neutrinos $ \nu_e,\nu_\mu,\nu_\tau$.

       Then we evaluate mass of right-handed neutrino by using seesaw relation (Eq.(5)). Oscillation experiments only gives neutrino mass square difference 
\begin{eqnarray*}
\Delta m_{12}^2&=&m_2^2-m_1^2   \\
\Delta m_{23}^2&=&m_3^2-m_2^2  
\end{eqnarray*}
which in our formalism can be rewritten as 
\begin{eqnarray*}
\Delta m_{12}^2&=&m_{\nu_\mu}^2-m_{\nu_e}^2  \\
\Delta m_{23}^2&=&m_{\nu_\tau}^2-m_{\nu_\mu}^2  
\end{eqnarray*}
 Thus we will use astronomical experimental data  (CMB, WMAP\cite{WMAP-7}, Supernova etc.) which gives the limit of sum of neutrino mass $ \sum m_\nu <0.58\ eV $ ( 95 \% CL ), to evaluate right-handed neutrino mass $ M_N $.  Assumed that neutrino mass is equal and then  $ m_\nu \approx 0.1eV$. The neutrino Dirac mass is comparable to the mass of corresponding charged lepton, then  
 \begin{eqnarray*}
 m_{eD}&\approx &O\left(m_{e^-}\right)\approx O\left(0.5MeV\right)\approx O\left({10}^5eV\right)\\
 m_{\mu D}&\approx& O\left(m_{\mu^-}\right)=O\left(105MeV\right)\approx O\left({10}^8eV\right)\\
 m_{\tau D}&\approx& O\left(m_{\tau^-}\right)=O\left(1776MeV\right)\approx O\left({10}^9eV\right)
 \end{eqnarray*}
 where $ m_{eD},m_{\mu D},m_{\tau D} $ are neutrino Dirac masses and $ m_{e^-},m_{\mu^-},m_{\tau^-} $ are charged lepton  ($e^-,\mu^-,\tau^-$) masses. Then the mass of right-handed neutrino according to seesaw relation is
 \begin{eqnarray}
M_{eN} &\approx& -\frac{M_{eD}^2}{m_\nu}\approx-\frac{\left(10^5\right)^2}{0.1}\sim-{10}^{11}eV=-0.1TeV  \\
M_{\mu N}&\approx&-\frac{M_{\mu D}^2}{m_\nu}=-\frac{\left(105\times{10}^6\right)^2}{0.1}\sim-{10}^{17}eV=-{10}^5TeV  \nonumber \\
\\
M_{\tau N}&\approx&-\frac{M_{\tau D}^2}{m_\nu}=-\frac{\left(1776\times{10}^6\right)^2}{0.1}\sim{10}^{19}eV=-{10}^7TeV . \nonumber \\
\end{eqnarray}

Then we will evaluate transition probability in seesaw mechanism using above right-handed neutrino masses. Seesaw mechanism requires neutrino mass eigenstate to be superposition of left-handed state $\nu_{L\alpha} $ and right-handed state $ N_R$, $-\frac{M_N}{m_D}\nu_{L\alpha}+N_R $, which means neutrino  jumping between $ \nu_{L\alpha} $ and $N_R$ during flight, appearing as left-handed state $ \nu_{L\alpha} $ by probability $  \frac{\left|\frac{M_N}{m_D}\right|^2}{\left|\frac{M_N}{m_D}\right|^2+1}\approx1 $, and appearing as right handed state $ N_R $ by probability $ \frac{1}{\left|\frac{M_N}{m_D}\right|^2+1}\approx\left|\frac{m_D}{M_N}\right|^2$. For example, the probability of electron neutrino appearing as right-handed state $ N_{eR} $ is
\begin{eqnarray}
 P\left(\nu_{eL}\rightarrow N_{eR}\right)=\left|\frac{m_{eD}}{M_{eN}}\right|^2\approx \left|\frac{10^5}{{10}^{11}}\right|^2=10^{-12},
 \end{eqnarray}
the probability of muon neutrino appearing as right-handed state $ N_{\mu R} $ is
\begin{eqnarray}
P\left(\nu_{\mu L}\rightarrow N_{\mu R}\right)=\left|\frac{m_{\mu D}}{M_{\mu N}}\right|^2\approx \left|\frac{10^8}{{10}^{17}}\right|^2\approx{10}^{-18},
\end{eqnarray}
the probability of tau neutrino appearing as right-handed state $ N_{\tau R} $ is 
\begin{eqnarray}
P\left(\nu_{\tau L}\rightarrow N_{\tau R}\right)=\left|\frac{m_{\tau D}}{M_{\tau N}}\right|^2\approx \left|\frac{10^9}{{10}^{19}}\right|^2={10}^{-20}.
\end{eqnarray}

\subsection{Apparent ‘fake’ neutrino mixture and neutrino oscillation}
 Right-handed neutrino mixture will induce neutrino apparent ‘fake’ mixture by seesaw mechanism as shown in Fig. 1, which can be expressed by Eq. (1) 
\begin{eqnarray*}
m_\nu
&\approx&-\frac{m_Dm_D^T}{M_N}=-\frac{m_Dm_D^T}{U_N^*M_N^d U_N^\dag}=-\frac{U_N^T m_Dm_D^TU_N}{M_N^d}\nonumber \\
&=&U^*\left(\begin{matrix}m_e&&\\&m_\mu&\\&&m_\tau\\\end{matrix}\right)U^\dag \\
&&\\
M_N^d&=&\left(\begin{matrix}M_{Ne}&&\\&M_{N\mu}&\\&&M_{N\tau}\\\end{matrix}\right)
\end{eqnarray*}
where neutrino mixing matrix $U$ is conjugate transpose matrix of right-handed neutrino mixing matrix $U_N$, $U=U_N^\dag$. As previously mentioned, neutrino mixing is not real mixture which is actually the mixture of right-handed neutrino, so we call it apparent mixture. Though neutrino mixing is 'fake' mixture in our formalism, it can still depict neutrino oscillation by oscillation probability formula Eq. (2)
\begin{eqnarray*}
P\left(\nu_\alpha\rightarrow\nu_\beta\right)&=&{sin}^2\left(2\theta_{ij}\right){sin}^2\left(\frac{t}{L_0}\right) \ ( \alpha, \beta= e,\mu, \tau)\\
L_0&=&\frac{4E}{\mathrm{\Delta}m_{ij}^2},   \nonumber   
\end{eqnarray*}
where mass square difference $\Delta m_{ij}^2 \ (i,j=e,\mu,\tau)$.
\begin{eqnarray}P\left(\nu_\alpha\rightarrow\nu_\beta\right)&=&{sin}^2\left(2\theta_{ij}\right){sin}^2\left(\frac{t\mathrm{\Delta}m_{ij}^2}{4E}\right) \nonumber \\
&=&{sin}^2\left(2\theta_{ij}\right){sin}^2\left(\frac{x\mathrm{\Delta}m_{ij}^2}{4E}\right)\\
&&( \alpha, \beta= e,\mu, \tau) \nonumber
\end{eqnarray}

Consider oscillation probability on the action distance of weak interaction $x\sim O\left(\frac{1}{M_Z}\right)=O\left(\frac{1}{90GeV}\right) $ 
\begin{eqnarray}
P\left(\nu_\alpha\rightarrow\nu_\beta\right)={sin}^2\left(2\theta_{ij}\right){sin}^2\left(\frac{\mathrm{\Delta}M_{ij}^2}{360GeV\times E}\right) \\
( \alpha, \beta= e,\mu, \tau).\nonumber
\end{eqnarray}
Given neutrino energy $ 1GeV $ and with the following good-fit data (which is adopted from PDG)
\begin{eqnarray*}
 \mathrm{\Delta}m_{21}^2&=&7.53\times{10}^{-5}eV^2\\
\mathrm{\Delta}m_{32}^2&=&2.41\times{10}^{-3}eV^2\\
{sin}^2\left(2\theta_{12}\right)&=&0.851\\
{sin}^2\left(2\theta_{23}\right)&=&0.991\\
{sin}^2\left(2\theta_{13}\right)&=&0.08
\end{eqnarray*}
 we can evaluate flavor conversion probability  on the action distance $ x$, which we take as probability at one point. The conversion probabilities on this distance $ x $ are 
\begin{eqnarray}P\left(\nu_e\rightarrow\nu_\mu\right)&=&{sin}^2\left(2\theta_{12}\right){sin}^2\left(\frac{\mathrm{\Delta}m_{12}^2}{360GeV\times E}\right) \nonumber \\
&=&0.851\times{sin}^2\left(\frac{7.53\times{10}^{-5}eV}{360GeV\times1GeV}\right) \nonumber\\
&=&3.72\times{10}^{-50} \\
P\left(\nu_e\rightarrow\nu_\tau\right)&=&{sin}^2\left(2\theta_{13}\right){sin}^2\left(\frac{\mathrm{\Delta}m_{31}^2}{360GeV\times E}\right)\nonumber \\
&=&0.08\times{sin}^2\left(\frac{2.41\times{10}^{-3}eV^2}{360GeV\times1GeV}\right)\nonumber\\
&=&3.59\times{10}^{-48} \\
P\left(\nu_\mu\rightarrow\nu_\tau\right)&=&{sin}^2\left(2\theta_{23}\right){sin}^2\left(\frac{\mathrm{\Delta}m_{32}^2}{360GeV\times E}\right)\nonumber\\
&=&0.991\times{sin}^2\left(\frac{2.41\times{10}^{-3}eV^2}{360GeV\times1GeV}\right)\nonumber\\
&=&4.44\times{10}^{-47}
\end{eqnarray}
        
        In our formalism, flavor conversion is induced by  chain of transition $  \nu_\alpha\rightarrow N_\alpha\rightarrow N_\beta\rightarrow\nu_\beta\ ( \alpha, \beta= e,\mu, \tau)$ and conversion probability at one point is 
\begin{eqnarray}
P\left(v_\alpha\rightarrow v_\beta\right)&=&P\left(v_\alpha\rightarrow N_\alpha\right)P\left(N_\alpha\rightarrow N_\beta\right)P\left(N_\beta\rightarrow\nu_\beta\right) \nonumber \\
( \alpha, \beta= e,\mu, \tau).&& 
\end{eqnarray}
For example, flavor conversion probability of $ \nu_e\rightarrow N_e\rightarrow N_\mu\rightarrow\nu_\mu $ at one point is
\begin{eqnarray}
 \begin{aligned}P\left( v_{e}\rightarrow v_\mu \right) =P\left( v_{e}\rightarrow N_e\right) P\left( N_e\rightarrow N_{\mu}\right) P\left( N_{\mu}\rightarrow \nu _{\mu}\right) \\
\sim 10^{-12}\times P\left( N_{e}\rightarrow N_ {\mu} \right) \times 1\approx 10^{-50}\\
\rightarrow P\left( N_{e}\rightarrow N_\mu \right) \sim 10^{-38}\end{aligned} \nonumber \\
\end{eqnarray}
For  $  \nu_e\rightarrow N_e\rightarrow N_\tau\rightarrow \nu_\tau$
\begin{eqnarray}
 \begin{aligned}P\left( v_{e}\rightarrow v_\tau \right) =P\left( v_{e}\rightarrow N_e\right) P\left( N_e\rightarrow N_{\tau}\right) P\left( N_{e}\rightarrow \nu _{\tau}\right) \\
\sim 10^{-12}\times P\left( N_{e}\rightarrow N_ {\tau} \right) \times 1\approx 10^{-48}\\
\rightarrow P\left( N_{e}\rightarrow N_\tau \right) \sim 10^{-36}\end{aligned}\nonumber \\
\end{eqnarray}
For   $ \nu_\mu\rightarrow N_\mu\rightarrow N_\tau\rightarrow\nu_\tau$
\begin{eqnarray}
\begin{aligned}P\left( v_{\mu}\rightarrow v_\tau \right) =P\left( v_{\mu}\rightarrow N_{\mu}\right) P\left( N_{\mu}\rightarrow N_{\tau}\right) P\left( N_{\tau}\rightarrow \nu _{\tau}\right) \\
\sim 10^{-18}\times P\left( N_{\mu}\rightarrow N_ {\tau} \right) \times 1 \approx 10^{-47}\\
\rightarrow P\left( N_{\mu}\rightarrow N_\tau \right) \sim 10^{-29}\end{aligned} \nonumber \\
\end{eqnarray}
For  $ \nu_\mu\rightarrow N_\mu\rightarrow N_e\rightarrow\nu_e$
\begin{eqnarray}
 \begin{aligned}P\left( v_{\mu}\rightarrow v_e \right) =P\left( v_{\mu}\rightarrow N_{\mu}\right) P\left( N_{\mu}\rightarrow N_{e}\right) P\left( N_{e}\rightarrow \nu _{e}\right) \\
\sim 10^{-18}\times P\left( N_{\mu}\rightarrow N_ e \right) \times 1 \approx 10^{-50}\\
\rightarrow P\left( N_{\mu}\rightarrow N_ e \right) \sim 10^{-32}\end{aligned} \nonumber\\
\end{eqnarray}
For  $ \nu_\tau\rightarrow N_\tau\rightarrow N_e\rightarrow\nu_e$
\begin{eqnarray}
\begin{aligned}P\left( v_\tau\rightarrow v_e \right) =P\left( v_{\tau}\rightarrow N_{\tau}\right) P\left( N_{\tau}\rightarrow N_{e}\right) P\left( N_{e}\rightarrow \nu _{e}\right) \\
\sim 10^{-20}\times P\left( N_{\tau}\rightarrow N_ e \right) \times 1 \approx 10^{-48}\\
\rightarrow P\left( N_{\tau}\rightarrow N_ e \right) \sim 10^{-28}\end{aligned}\nonumber \\
\end{eqnarray}  
For   $  \nu_\tau\rightarrow N_\tau\rightarrow N_\mu\rightarrow\nu_\mu$
\begin{eqnarray}
\begin{aligned}P\left( v_{\tau}\rightarrow v_\mu \right) =P\left( v_{\tau}\rightarrow N_{\tau}\right) P\left( N_{\tau}\rightarrow N_{\mu}\right) P\left( N_{\mu}\rightarrow \nu _{\mu}\right) \\
\sim 10^{-20}\times P\left( N_{\tau}\rightarrow N_ {\mu} \right) \times 1 \approx 10^{-47}\\
\rightarrow P\left( N_{\tau}\rightarrow N_\mu \right) \sim 10^{-27}\end{aligned}\nonumber\\
\end{eqnarray}

\subsection{Flavor conversion probability of right-handed neutrino and flavor flip interaction between right-handed neutrinos}

Right-handed neutrino flavor conversion due to right-handed neutrino mixture  can be considered unknown flavor flip interaction between right-handed neutrinos as shown in Fig.~2.
\begin{figure}[!hpb]
\includegraphics[width=0.4\textwidth]{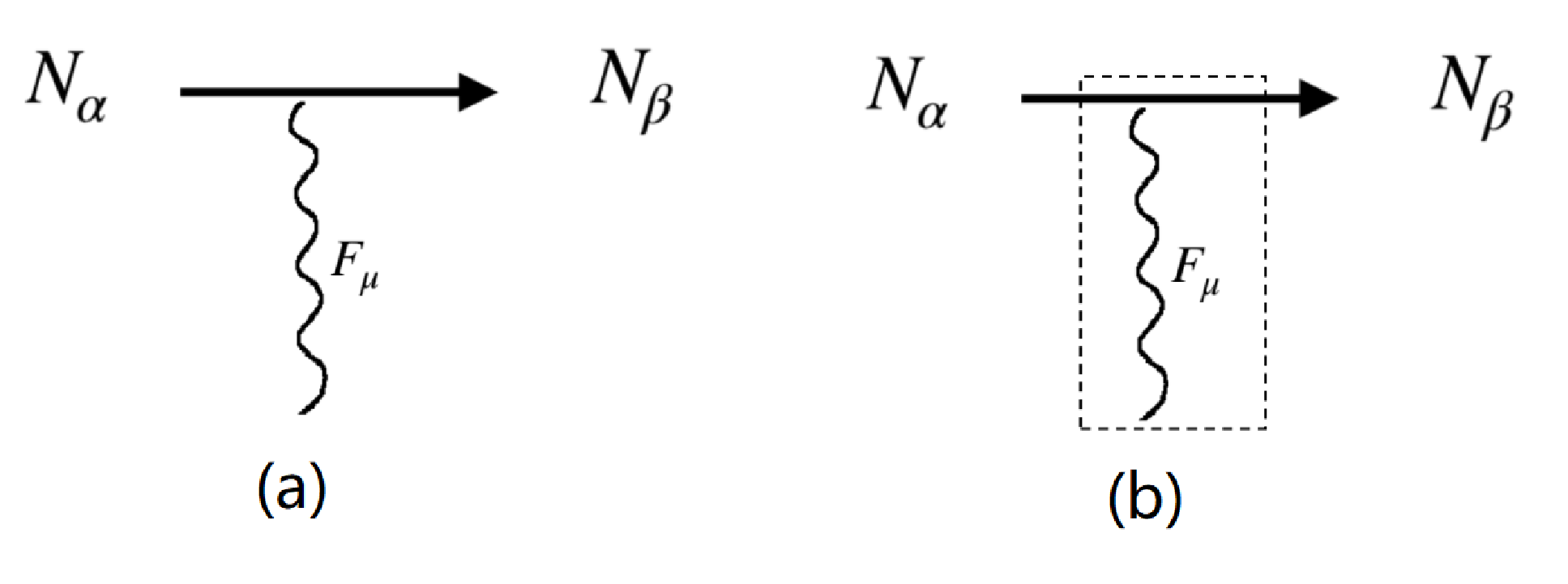}
\caption{ Right-handed neutrino flavor conversion due to right-handed neutrino mixture (b) can be considered as unknown flavor flip interaction between right-handed neutrino (assumed intermediated by Boson F) (a) ($ \alpha, \beta= e,\mu, \tau$). 
}
\end{figure}
We assume the flavor flip interaction between right-handed neutrino is intermediated by boson F just like boson W,Z intermediating weak interaction. Then the strength of this interaction can be crudely evaluated by computing the amplification of Fig.~3.
\begin{figure}[!hpb]
\includegraphics[width=0.2\textwidth]{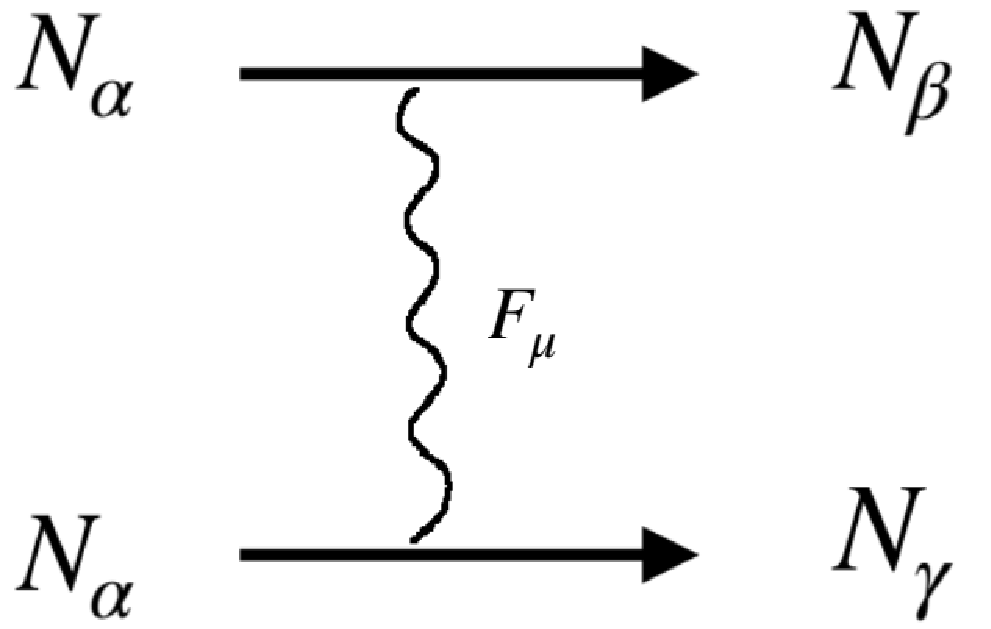}
\caption{ Flavor flip interaction between right-handed neutrino intermediated by boson F ($ \alpha, \beta,\gamma= e,\mu, \tau$).}
\end{figure}
\begin{eqnarray}
P\left(N_\alpha\rightarrow N_\beta\right)\sim\left|M\right|^2\sim\frac{32g_F^4}{M_F^4}\times M_N^4 \\
( \alpha, \beta= e,\mu, \tau) \nonumber
\end{eqnarray}
We choose arbitrarily $ N_e\rightarrow N_\mu $ mode to evaluate the interaction strength by Eq. (24)
\begin{eqnarray}
P\left(N_e\rightarrow N_\mu\right)&\approx&\left|M\right|^2\approx\frac{32g_F^4}{M_F^4} \times M_N^4 \nonumber \\
&\sim&\frac{32g_F^4}{M_F^4}\times M_{N_e}^2\times M_{N_\mu}^2  \nonumber    \\
&\sim&{10}^{-38} 
\end{eqnarray}
Then we have the relation 
\begin{eqnarray}
\frac{g_F^2}{M_F^2}\sim\frac{{10}^{-29}}{GeV^2}
\end{eqnarray}
The weak interaction strength
\begin{eqnarray}
 \frac{g_W^2}{M_W^2}=\frac{G_F}{\sqrt2}=\frac{{10}^{-5}}{GeV^2} 
\end{eqnarray} 
and then
\begin{eqnarray}
\frac{g_F^2}{M_F^2}\approx\frac{{10}^{-29}}{GeV^2}\approx{10}^{-24}G_F 
\end{eqnarray} 
Thus, the assumed flavor flip interaction $ N_\alpha\leftrightarrow N_\beta $ is very week interaction. Assumed $ g_F\sim g_z\sim1$ (referring to PDG), we can evaluate $ M_F\sim{10}^{14}GeV=10^{11}TeV $ which is more heavier than right-handed neutrino mass. For the lepton number of right-handed neutrino is zero, flavor flip interaction between right-handed neutrinos is allowed in the formalism of the standard model.

Intermediator F is heavier than right-handed neutrino, and is more liable to decay to heavier right-handed neutrino to accomplish flavor conversion. Thus the right-handed neutrino flavor conversion probability according to experimental data Eq. (18)$\sim$(23) can be well explained by flavor flip interaction between right-handed neutrinos.

\section{	 Conclusions}
This paper proposed a new formalism in which  neutrino flavor conversion is induced by right-handed neutrino mixture via seesaw mechanism which leads to apparent 'fake' neutrino mixture which 'generates' so-called 'neutrino oscillation'.

Neutrino mass eigenstate is superposition of left-handed state and right-handed state $ -\frac{M_N}{m_D}\nu_{L\alpha}+N_R \ (\alpha=e,\mu,\tau)  $   which is consistent with  neutrino flavor, regardless of right-handed neutrino flavor mixture. This apparent 'fake' neutrino mixture can still depict neutrino oscillation, except that mass square difference in oscillation probability formula $\Delta m_{12}^2=m_\mu^2-m_e^2, \ \Delta m_{23}^2=m_\tau^2-m_\mu^2  $. And $\delta_{CP}$ appearing in neutrino mixture is also the CP violation phase of right-handed neutrino mixture.
      
      We evaluate right-handed neutrino mass according to seesaw mechanism, with which we calculate right-handed neutrino flavor conversion probability according to our formalism. We find that it can be explained well by boson (denoted as F) intermediating flavor flip interaction between right-handed neutrino which  strength $  \frac{g_F^2}{M_F^2}\approx\frac{{10}^{-29}}{GeV^2}={10}^{-24}G_F $.

\begin{acknowledgments}
I wish to acknowledge the support of Sichuan University of Arts and Science,  acknowledge Prof. Guang-Ping Chen for his useful suggestion, and acknowledge Doctor Zhi-Hao Yang for his useful suggestion.
\end{acknowledgments}
       
\nocite{*}
\bibliography{neutrino}

\end{document}